\documentclass[aps,prd,twocolumn,showpacs,showkeys,nofootinbib]{revtex4}
\pdfoutput=1
\usepackage{epsfig}
\usepackage{amsmath}
\usepackage{graphicx}
\newcommand{\beq}{\begin{equation}}
\newcommand{\eeq}{\end{equation}}
\newcommand{\bqa}{\begin{eqnarray}}
\newcommand{\eqa}{\end{eqnarray}}

\begin{document}

\title{Hidden charm octet tetraquarks from a diquark-antidiquark model }
\author{Ruilin Zhu$^{1,2}$  }
\affiliation{
$^1$ Department of Physics and Institute of Theoretical Physics,
Nanjing Normal University, Nanjing, Jiangsu 210023, China\\
$^2$ INPAC, Shanghai Key Laboratory for Particle Physics and Cosmology, Department of Physics and Astronomy, Shanghai Jiao Tong University, Shanghai, 200240,   China\\
 }

\begin{abstract}
Four exotic charmonium-like states, i.e. $X(4140)$, $X(4274)$, $X(4500)$, and $X(4700)$, have been observed very recently by LHCb Collaboration  in the decay process  $B^+\to  J/\psi \phi K^+$ using  the 3${\rm fb}^{-1}$ data of $p\bar p$ collision at $\sqrt s= 7$ and $8$ TeV. In this paper, we investigate systematically  the properties of hidden charm tetraquark states.   The hidden charm tetraquarks form an octet configuration and a singlet configuration according to flavor $SU(3)$ symmetry. Based on a diquark-antidiquark model,  the hidden charm tetraquarks spectra are given.
The previous XYZ exotic states altogether with the newly ones $X(4140)$, $X(4274)$, $X(4500)$, and $X(4700)$, can be  well classified into  certain representations. The spin-parities and masses of the XYZ are predicted, most of which are in agreement with the data. We particularly find that $Z_c(4430)$ may be treated as the first radial excitation of  $Z_c(3900)$, while the $Y(1^{--})$  states can be obtained by the first orbital excitation of the $X/Z$ states. Besides, we  calculate the decay widths of the hidden charm  tetraquarks into two charmed mesons. This work gives a general tetraquark description for the XYZ states, which is helpful to uncover their inner structures.

\pacs{14.40.Rt, 12.39.Mk,  12.40.Yx}
\keywords{Exotic state,  tetraquark, diquark }

\end{abstract}

\maketitle

\section{Introduction}

The exotic tetraquarks consisting of two quarks and two antiquarks are not ruled out in
Quantum Chromodynamics (QCD). As the argument by Steven Weinberg, the Fourier transform of
the connected part of the three-point  Green function  of the tetraquark and two ordinary mesons can
contain a tetraquark pole, which gives a decay rate into two light ordinary mesons proportional to
$1/N_c$ with color SU(3) group parameter $N_c$ ~\cite{Weinberg:2013cfa}.

The history of the XYZ exotic states can be traced back to the discovery of $X(3872)$ by the Belle Collaboration in 2003~\cite{Choi:2003ue}, which lies above the open charm threshold but has a very narrow  decay width ($\Gamma<1.2$MeV). Two years later, $Y(4260)$ with spin-parity quantum
numbers $1^{--}$ was discovered in
the process $e^+e^-\to \gamma (\pi^+\pi^- J/\psi)$ by the BaBar Collaboration\cite{Aubert:2005rm}, which can decay
 to two charmed mesons. In 2007, the Belle
Collaboration observed  a charged hidden charm tetraquark state, i.e. $Z^+(4430)$~\cite{Choi:2007wga}.
Through $Y(4260) \to \pi^-\pi^+J/\psi$, the BESIII Collaboration discovered $Z_c^+(3900)$ in 2013~\cite{Ablikim:2013mio}, which can decay into
$J/\psi\pi^+$ through strong interactions, which implies that it shall be  a  meson with quark contents $c\bar{c}u\bar{d}$. Besides these remarkable exotic states, other XYZ mesons have been observed by different experiments over the past twelve years. The list of XYZ mesons has grown to about two dozen up to date.

Very recently,  the LHCb collaboration has reported four $J/\psi \phi$  structures, i.e. $X(4140)$, $X(4274)$, $X(4500)$, and $X(4700)$ in the process  $B^+\to  J/\psi \phi K^+$ using  the 3${\rm fb}^{-1}$ data of $p\bar p$ collision at $\sqrt s= 7$ and $8$ TeV~\cite{Aaij:2016iza,Aaij:2016nsc}. Each structure has the
significance over 5 standard deviations.  Their masses  and  decay widths have been determined as~\cite{Aaij:2016iza}
\begin{eqnarray}
 M_{X(4140)}&=& (4146.5\pm4.5_{-2.8}^{+4.6}){\rm MeV}, \nonumber\\
  \Gamma_{X(4140)}&=& (83\pm 21_{-14}^{+21}){\rm MeV},\nonumber\\
 M_{X(4274)}&=& (4273.3\pm8.3_{-3.6}^{+17.2}){\rm MeV}, \nonumber\\
  \Gamma_{X(4274)}&=& (56\pm 11_{-11}^{+8}){\rm MeV},\nonumber\\
 M_{X(4500)}&=& (4506\pm11_{-15}^{+12}){\rm MeV}, \nonumber\\
  \Gamma_{X(4500)}&=& (92\pm 21_{-20}^{+21}){\rm MeV},\nonumber\\
   M_{X(4700)}&=& (4704\pm10_{-24}^{+14}){\rm MeV}, \nonumber\\
  \Gamma_{X(4700)}&=& (120\pm 31_{-33}^{+42}){\rm MeV}.\nonumber
\end{eqnarray}
The above results are obtained from naive $J/\psi \phi$ mass fits with a Breit-Wigner parametrization. The first errors are statistical, and the second errors are systematic. These four $J/\psi \phi$  structures have attracted many theorists' attention~\cite{Chen:2016oma,Wang:2016gxp,Wang:2016tzr,Ali:2016dkf,Liu:2016onn,Maiani:2016Jul}.

Many new interesting structures were discovered in the mass region of heavy
quarkonium, which call for a general approach  to  understand  and predict these XYZ states.
A few general schemes are proposed, such as Born-Oppenheimer approximation~\cite{Braaten:2014qka}
 and compact tetraquark model~\cite{Brodsky:2014xia}, but the spectra are not given.
For a general review of the $XYZ$ states, see Refs.~\cite{Brambilla:2010cs,Agashe:2014kda,Esposito:2014rxa,Chen:2016qju}.
 In the past decades,  hidden charm and bottom tetraquark states  have been investigated in Refs.~\cite{Chen:2004dy,Cheng:2003kg,Maiani:2004vq,Ali:2009pi,Ali:2009es,Ali:2010pq,Chen:2010ze,Ali:2011ug,Ali:2014dva,Liu:2014dla,
Maiani:2014aja,Ma:2014zva,Wang:2016mmg,Lebed:2016yvr,Wang:2016tsi,He:2016xvd,He:2016yhd}.

In this paper, we investigate systematically the mass spectra and decay widths of hidden charm tetraquarks in a diquark-antidiquark model.
The hidden charm tetraquarks can be  organized into octet and singlet representations in flavor SU(3) symmetry.
We interpret the XYZ exotic states  as tetraquarks, and study the possible inner relations among different XYZ states.
According to these  studies,  we want to obtain a general description for the XYZ states altogether.

The paper is organized as the following. In Sec.~\ref{II}, we classify the hidden charm tetraquarks into octet and singlet representations in  flavor SU(3) symmetry.
 In Sec.~\ref{III}, hidden charm octet and singlet tetraquarks spectra are calculated. The XYZ exotic states
 are interpreted as certain tetraquarks. In Sec.~\ref{IV}, we study
 the two-body charmed mesons decays of tetraquarks. We summarize and conclude in the end.

\section{Hidden charm octet and singlet tetraquarks in SU(3)\label{II}}

Hidden charm tetraquark states $X/Y/Z \sim [cq']\bar{q}\bar{c}$ with two light quarks, $q$ and $q'$, can be conventionally classified by flavor $SU(3)$ symmetry.
Under the flavor $SU(3)$ symmetry, the three light quarks, $(u,\; d,\; s)$ form a triplet ${\bf 3}$ representation and the charm quark $c$ is a singlet~\cite{Zeppenfeld:1980ex,Chau:1990ay,Gronau:1994rj}. Tetraquark states formed by two light quarks and a charm quark-antiquark pair can have the following irreducible representations
\begin{eqnarray}
{\bf 3} \otimes  {\bf\bar 3} = {\bf8}\oplus {\bf1}\;.
\end{eqnarray}

Hidden charm tetraquarks form an octet and a singlet.  Different tetraquarks can
be rearranged together. The octet tetraquarks can be expressed as
\begin{eqnarray}
E^i_j=\begin{pmatrix}
 \frac{Z_c^0}{\sqrt{2}}+\frac{X}{\sqrt{6}}  &Z_c^+ &Z_{cs}^+\\
 Z_c^-&-\frac{Z_c^0}{\sqrt{2}}+\frac{X}{\sqrt{6}}&{Z_{cs}^0}\\
 Z_{cs}^-&\bar Z_{cs}^0 &-2\frac{X}{\sqrt{6}}
 \end{pmatrix},
\end{eqnarray}
where we adopt the conventional notation for the $Z_c$ Isospin triplet, i.e. $Z_c^0$ and $Z_c^\pm$.
Similarly, $Z_{cs}$ is introduced to denote two Isospin doublets, i.e. $Z_{cs}^0$, $\bar Z_{cs}^0$, and $Z_{cs}^\pm$.
$X$ denotes an Isospin singlet here, which should be distinguished  from the notation for a general exotic state $X$.

The orbitally excited tetraquarks also form an octet and a singlet. Considering the first orbital excitation between diquark and antidiquark in
the  tetraquark  with $L=1$,
the possible octet tetraquark states with spin-parity $1^-$ can be written as
\begin{eqnarray}
{E^*(1^-)}^i_j=\begin{pmatrix}
 \frac{Z_c^{*0}}{\sqrt{2}}+\frac{X^*}{\sqrt{6}}  &Z_c^{*+} &Z_{cs}^{*+}\\
{Z_c^{*-}}&-\frac{Z_c^{*0}}{\sqrt{2}}+\frac{X^*}{\sqrt{6}}&Z_{cs}^{*0}\\
{Z_{cs}^{*-}}& {\bar{Z}}_{cs}^{*0} &-2\frac{X^*}{\sqrt{6}}
 \end{pmatrix},
\end{eqnarray}
where the neutral tetraquarks $Z_c^{*0}$ and $X^*$ can have the definite charge-parity,
thus their $J^{PC}$ quantum numbers can be $1^{--}$ or $1^{-+}$.

The SU(3) flavor singlet can be denoted as $X'$ ($X'^*$). Unlike the $\eta '$ where the gluonium content may exist~\cite{Muta:1999tc,Ali:2000ci}, we ignore the mixing effects from $[c\bar{c}gg]$ which is
  a state with charm quark-antiquark pair and two gluons for simplification~\cite{Zhu:2015qoa,Zhu:2015jha}. Other orbitally excited tetraquarks with different spin-parity will have
a similar group representation decomposition.

The flavor components of tetraquarks  in flavor SU(3) symmetry can be  explicitly given as below
\begin{eqnarray}
 &&Z_c^0= \frac{1}{\sqrt 2} (u\bar u-d \bar d)c\bar c, \;\;\;  X = \frac{1}{\sqrt 6} (u\bar u+d \bar d -2s \bar s)c\bar c, \nonumber\\
 &&Z_c^+=  u\bar d c\bar c, \;\;\; Z_c^-=  d\bar u c\bar c, \;\;\; Z_{cs}^+ = u\bar s c\bar c, \nonumber\\
  &&Z_{cs}^- = s\bar u c\bar c, \;\;\; Z_{cs}^0=  d\bar s c\bar c, \;\;\; {\bar Z}_{cs}^0 = s\bar d c\bar c, \nonumber\\
    &&X' = \frac{1}{\sqrt 3} (u\bar u+d \bar d +s \bar s)c\bar c.
\end{eqnarray}
The above classification for hidden charm tetraquarks into an octet and a singlet  is much similar to
the classification for conventional light mesons. Current experimental data also do support the evidence that the tetraquark is not alone, but has a lot of companions. These companions shall be organized into these two representations according to flavor SU(3) symmetry.

\section{Hidden charm octet and singlet tetraquarks spectra\label{III}}

The general QCD confining potential for the multiquarks are~\cite{DeRujula:1975ge}
\begin{equation}\label{potential}
V(\vec{r}_i) = L(\vec{r}_1,\vec{r}_2,\ldots)+\sum_{i>j}I\, \alpha_s S_{ij}\ ,
\end{equation}
where $L(\vec{r}_i)$ denotes the universal binding interaction of quarks, $S_{ij}$ are two-body Coulomb and chromomagnetic interactions, and $I=-\frac{4}{3}$ and $-\frac{2}{3}$ represent the strength coefficient in the quark-antiquark and quark-quark cases, respectively. According to color $SU(3)$ symmetry, a quark-antiquark pair can be
 represented either by a singlet or octet in the decomposition of   $\mathbf{3}\otimes \mathbf{\bar 3}=\mathbf{1}\oplus \mathbf{8}$, and a diquark can be represented either by an antitriplet or sextet in the decomposition of $\mathbf{3}\otimes \mathbf{3}=\mathbf{\bar{3}}\oplus \mathbf{6}$. In the one-gluon-exchange model,
the binding of the $q_1\bar{q}_2$ or $q_1q_2$ system depends solely on the quadratic Casimir $C_2(R)$ of the product color representation
R to which the quarks couple according to the discriminator
$I=\frac{1}{2}(C_2(R)-C_2(R_1)-C_2(R_2))$, where $R_i$ denotes the color representations of two quarks \cite{Brodsky:2014xia}. One can immediately obtain the strength
coefficients $I=\frac{1}{6}(-8,-4,+2,+1)$ for $R = (\mathbf{1},\mathbf{\bar{3}},\mathbf{6},\mathbf{8})$, respectively. The color singlet of  a quark-antiquark pair
and the color antitriplet of the diquark are the two kinds of attractive representation.

The effective Hamiltonian includes three categories of interactions: spin-spin interactions of quarks in the diquark and antidiquark, and  between them;  spin-orbital interactions;  orbit-orbital interactions. The explicit forms can be written as~\cite{Maiani:2004vq}:
\begin{eqnarray}
 H&=&m_{\delta}+m_{\delta^\prime}+H^{\delta}_{SS} + H^{\bar{\delta^\prime}}_{SS}+H^{\delta\bar{\delta^\prime}}_{SS} + H_{SL}+H_{LL},\nonumber\\
 \label{eq:definition-hamiltonian}
\end{eqnarray}
with the interaction
\begin{eqnarray}
 H^\delta_{SS}&=&2(\kappa_{c q'})_{\bar{3}}(\mathbf{S}_c\cdot \mathbf{S}_{q'}),\nonumber\\
 H^{\bar{\delta^\prime}}_{SS}&=&2(\kappa_{ cq})_{\bar{3}}(\mathbf{S}_{\bar{c}}\cdot \mathbf{S}_{\bar{q}}), \nonumber\\
 H^{\delta\bar{\delta^\prime}}_{SS} &=&2\kappa_{q'\bar{q}}(\mathbf{S}_{q'}\cdot \mathbf{S}_{\bar{q}}) +2\kappa_{c \bar{q}}(\mathbf{S}_{c}\cdot \mathbf{S}_{\bar{q}})\nonumber\\&&+2\kappa_{c\bar{q}'} (\mathbf{S}_{q'}\cdot \mathbf{S}_{\bar{c}})+ 2\kappa_{c\bar{c}}(\mathbf{S}_{c}\cdot \mathbf{S}_{\bar{c}}),
\nonumber\\
 H_{SL}&=&2 A_\delta (\mathbf{S}_\delta \cdot \mathbf{L}) +2 A_{\bar{\delta^\prime}}(\mathbf{S}_{\bar{\delta^\prime}}\cdot \mathbf{L}),\nonumber\\
 H_{LL}&=&B_{\delta\bar{\delta^\prime}} \frac{L(L+1)}{2}\ .
\label{eq:definition-hamiltonian2}
\end{eqnarray}
where  $m_\delta$ and $m_{\delta^\prime}$ are the constituent masses of the diquark $[cq^\prime]$ and the antidiquark $[\bar{q}\bar{c}]$, respectively.  $ H^\delta_{SS}$ and $H^{\bar{\delta^\prime}}_{SS}$ denotes the spin-spin interaction inside the diquark and antidiquark, respectively. $H^{\delta\bar{\delta^\prime}}_{SS}$ describes  the spin-spin interaction of quarks between diquark and antidiquark. $H_{SL}$ and $H_{LL}$ are the spin-orbital and purely orbital terms.  $\mathbf{S}_{\delta}$ and $\mathbf{S}_{\bar{\delta^\prime}}$ correspond to the spin operators  of diquark and antidiquark, respectively. The spin operators  of light and  heavy quarks  are given by  $\mathbf{S}_{q^{(\prime)}}$ and $\mathbf{S}_{c}$, respectively.  $\mathbf{L}$ denotes  the orbital angular momentum operator.  The coefficients $\kappa_{q_1\bar{q}_2}$ and $(\kappa_{q_1 q_2})_{\bar{3}}$ are the spin-spin couplings for a quark-antiquark pair and diquark in color antitriplet ${\bf\bar 3}$, respectively; $A_{\delta(\bar{\delta^\prime})}$ and $B_{\delta\bar{\delta^\prime}}$ denote spin-orbit and orbit-orbit couplings, respectively.

The lowest-lying  tetraquark states have vanishing orbital angular momenta, i.e. $L=0$.  There are two possible tetraquark configurations  with the spin-parity $J^P=0^+$, i.e.,
\begin{eqnarray}
|0_\delta,0_{\bar{\delta^\prime}},0_J\rangle&=&\frac{1}{2}
\big[(\uparrow)_c(\downarrow)_{q^\prime}-(\downarrow)_c(\uparrow)_{q^\prime} \big](\uparrow)_{\bar{q}}(\downarrow)_{\bar{c}}
\nonumber\\
&&-\frac{1}{2}
\big[(\uparrow)_c(\downarrow)_{q^\prime}-(\downarrow)_c(\uparrow)_{q^\prime} \big](\downarrow)_{\bar{q}}(\uparrow)_{\bar{c}},\nonumber\\
|1_\delta,1_{\bar{\delta^\prime}},0_J\rangle&=&\frac{1}{\sqrt{3}}
\big\{(\uparrow)_c(\uparrow)_{q^\prime}(\downarrow)_{\bar{q}}(\downarrow)_{\bar{c}}
+(\downarrow)_c(\downarrow)_{q^\prime}(\uparrow)_{\bar{q}}(\uparrow)_{\bar{c}} \nonumber\\&& -\frac{1}{2}
\big[(\uparrow)_c(\downarrow)_{q^\prime}+(\downarrow)_c(\uparrow)_{q^\prime} \big](\uparrow)_{\bar{q}}(\downarrow)_{\bar{c}}
\nonumber\\&& -\frac{1}{2}
\big[(\uparrow)_c(\downarrow)_{q^\prime}+(\downarrow)_c(\uparrow)_{q^\prime} \big](\downarrow)_{\bar{q}}(\uparrow)_{\bar{c}}\big\}.
 \label{eq:definition-states0+}
\end{eqnarray}
where $|S_\delta,S_{\bar{\delta^\prime}},S_J\rangle $ stands for the tetraquark; the $S_\delta$ and $S_{\bar{\delta^\prime}}$ denote  the spin  of diquark $[cq^\prime]$ and antidiquark $[\bar{q}\bar{c}]$, respectively, while the $S_J$ denotes the total angular momentum of the tetraquark.

The base vectors defined in Eq.~(\ref{eq:definition-states0+}) are not the eigen-vectors of the effective Hamiltonian in Eq.~(\ref{eq:definition-hamiltonian}), thus the mass matrix is not diagonal. The splitting mass matrix for the $J^P=0^+$ tetraquarks is
\begin{eqnarray}
\Delta M (0^+)=\left(
\begin{array}{cc}
 -\frac{3}{2} ((\kappa_{c{q^\prime}})_{\bar{3}}+(\kappa_{c q})_{\bar{3}}) & h_1\\
 \frac{\sqrt{3}}{2}  (\kappa_{c \bar{q'}}+\kappa_{c\bar{q}}-\kappa_{q^\prime \bar{q}}-\kappa_{c\bar{c}}) & h_2
\end{array} \right).
\end{eqnarray}
with
\begin{eqnarray}
h_1&= &\frac{\sqrt{3}}{2}  (\kappa_{c \bar{q'}}+\kappa_{c\bar{q}}-\kappa_{q^\prime \bar{q}}-\kappa_{c\bar{c}}) ,\nonumber\\
h_2&= &\frac{1}{2} ((\kappa_{c{q^\prime}})_{\bar{3}}+(\kappa_{c q})_{\bar{3}}-2 \kappa_{q^\prime \bar{q}}-2 \kappa_{c \bar{q'}}-2 \kappa_{q\bar{q}}-2 \kappa_{c\bar{c}}).\nonumber\\
\end{eqnarray}
and the mass matrix is given as
\begin{eqnarray}
 M(J^P)= m_{\delta}+m_{\delta'}+\Delta M(J^P).
\end{eqnarray}
Diagonalizing  the above mass matrix, one can easily obtain two different eigenvalues of masses.

For $J^P=2^+$, there is only one tetraquark configuration
\begin{eqnarray}
|1_\delta,1_{\bar{\delta^\prime}},2_J\rangle&=&
(\uparrow)_c(\uparrow)_{q^\prime}(\uparrow)_{\bar{q}}(\uparrow)_{\bar{c}},
\label{eq:definition-states2+}
\end{eqnarray}
with the mass
\begin{eqnarray}
 M(2^+)&=&m_\delta+m_{\delta^\prime}+\frac{1}{2}\left( (\kappa_{c{q^\prime}})_{\bar{3}}+(\kappa_{c q})_{\bar{3}}\right)\nonumber\\&&+\frac{1}{2}\left(\kappa_{c\bar{q}}+\kappa_{c\bar{c}}+\kappa_{q^\prime \bar{q}}+\kappa_{c\bar{q'}}\right).
\end{eqnarray}

As for  $J^P=1^+$, there are three possible tetraquark configurations, i.e.,
\begin{eqnarray}
|0_\delta,1_{\bar{\delta^\prime}},1_J\rangle&=&\frac{1}{\sqrt{2}}
\big[(\uparrow)_c(\downarrow)_{q^\prime}-(\downarrow)_c(\uparrow)_{q^\prime} \big](\uparrow)_{\bar{q}}(\uparrow)_{\bar{c}}
 ,\nonumber\\
 |1_\delta,0_{\bar{\delta^\prime}},1_J\rangle&=&\frac{1}{\sqrt{2}}
(\uparrow)_c(\uparrow)_{q^\prime}\big[(\uparrow)_{\bar{q}}(\downarrow)_{\bar{c}}-(\downarrow)_{\bar{q}}
(\uparrow)_{\bar{c}}\big]
 ,\nonumber\\
|1_\delta,1_{\bar{\delta^\prime}},1_J\rangle&=&\frac{1}{2}
\big\{(\uparrow)_c(\uparrow)_{q^\prime}\big[(\uparrow)_{\bar{q}}(\downarrow)_{\bar{c}}+(\downarrow)_{\bar{q}}
(\uparrow)_{\bar{c}}\big]\nonumber\\&&-\big[(\uparrow)_c(\downarrow)_{q^\prime}+(\downarrow)_c(\uparrow)_{q^\prime} \big](\uparrow)_{\bar{q}}(\uparrow)_{\bar{c}}\big\}.
 \label{eq:definition-states1+}
\end{eqnarray}
Notice that  the tetraquarks with the quark content $[cq^\prime][\bar{q}\bar{c}]$ where $q^\prime\neq q$,  do not have any definite charge parity and thus the above three $1^+$ states can mix with each other.

Using the base vectors   defined in Eq. (\ref{eq:definition-states1+}), one can obtain  the mass splitting matrix $\Delta M$ for $J^P=1^+$

\begin{widetext}
\begin{eqnarray}
\Delta M=\left(
\begin{array}{ccc}
 \frac{1}{2} ((\kappa_{c q})_{\bar{3}}-3 (\kappa_{c{q^\prime}})_{\bar{3}}) & \frac{1}{2} (\kappa_{c \bar{q'}}-\kappa_{q^\prime \bar{q}}-\kappa_{c\bar{c}}+\kappa_{c\bar{q}}) & \frac{\sqrt{2}}{2}(\kappa_{c\bar{c}}-\kappa_{c\bar{q'}}-\kappa_{q^\prime \bar{q}}+\kappa_{c\bar{q}}) \\
 \frac{1}{2} (\kappa_{c\bar{q'}}-\kappa_{q^\prime \bar{q}}-\kappa_{c\bar{c}}+\kappa_{c\bar{q}}) & \frac{1}{2} ((\kappa_{c{q^\prime}})_{\bar{3}}-3 (\kappa_{c q})_{\bar{3}}) & \frac{\sqrt{2}}{2}(\kappa_{c\bar{q'}}-\kappa_{q^\prime \bar{q}}+\kappa_{c\bar{c}}-\kappa_{c\bar{q}}) \\
 \frac{\sqrt{2}}{2}(\kappa_{c\bar{c}}-\kappa_{c \bar{q'}}-\kappa_{q^\prime \bar{q}}+\kappa_{c\bar{q}}) & \frac{\sqrt{2}}{2}(\kappa_{c \bar{q'}}-\kappa_{q^\prime \bar{q}}+\kappa_{ c\bar{c}}-\kappa_{c\bar{q}}) & \frac{1}{2}
   ((\kappa_{c{q^\prime}})_{\bar{3}}+(\kappa_{c q})_{\bar{3}}-\kappa_{c \bar{q'}}-\kappa_{q^\prime \bar{q}}-\kappa_{ c\bar{c}}-\kappa_{c\bar{q}})
\end{array}
\right),\nonumber\\
\end{eqnarray}
\end{widetext}

In flavor SU(3) symmetry, all hidden charm tetraquark states with vanishing orbital angular momenta
will have the identical mass.  Considering the strange quark is different from the up and down quarks,
we will obtain the hidden charm tetraquark masses, which are reasonable to explain the experimental data. In the paper,
we adopt the  quark masses as $m_q=305\mathrm{MeV}, m_s=490\mathrm{MeV},m_c=1.670\mathrm{GeV}$~\cite{Maiani:2004vq,Zhu:2015bba,Wang:2016tsi}.
For the charmed diquark, we use $m_{cq}=1.932${GeV}~\cite{Zhu:2015bba}.
The  diquark mass $m_{cs}$ is estimated by the relation $m_{cs}\simeq m_{c}+m_{s}$.
The spin-spin couplings are~\cite{Maiani:2004vq,Zhu:2015bba}: $(\kappa _{cq})_{\bar 3}=22$MeV, $(\kappa _{cs})_{\bar 3}=25$MeV, $(\kappa _{q\bar{q}})_{0}=315$MeV, $(\kappa _{s\bar{q}})_{0}=195$MeV, $(\kappa _{s\bar{s}})_{0}=121$MeV, $(\kappa _{c\bar{q}})_{0}=70$MeV and $(\kappa _{c\bar{s}})_{0}=72$MeV. The relation $\kappa _{ij}=\frac{1}{4}(\kappa _{ij})_{0}$ for the quark-antiquark state is employed, which is derived from one gluon exchange model. The spin-orbit
coupling $A_{\delta}$ is  estimated as $30~\mbox{MeV}\ \mbox{and}\ 5~\mbox{MeV}$ for
$c$ and $b$ quarks, respectively; and the orbit-orbit coupling
$B_{\delta \bar{\delta}'}$  is  estimated as  $278~\mbox{MeV}$ and $408~\mbox{MeV}$ for $c$ and $b$ quarks,
respectively~\cite{Maiani:2004vq,Zhu:2015bba}. For the radial excitation, $500\sim600$MeV radial splitting is estimated reasonably from $\psi(2S)$ to $J/\psi$.

The spin of tetraquark states can be 0, 1 and 2, if there is no orbital excitation. The tetraquarks belonging to the same Isospin representation will have the identical mass due to Isospin
symmetry. Their masses are estimated to be:
\begin{align}\label{mass1}
 m(Z_c)&= \left\{ \begin{array} {ll}3.72 {\rm GeV}, 3.83 {\rm GeV} , & J^P=0^+ , \\ 3.75 {\rm GeV} ,3.87 {\rm GeV} ,3.88 {\rm GeV} , & J^P=1^+ ,
 \\ 3.94 {\rm GeV}, & J^P=2^+ ,\end{array} \right.\\
 m(Z_{cs}) &= \left\{ \begin{array} {ll}4.00 {\rm GeV}, 4.04 {\rm GeV} , & J^P=0^+ , \\ 4.03 {\rm GeV} ,4.08 {\rm GeV} ,4.09 {\rm GeV} , & J^P=1^+ ,
 \\ 4.17 {\rm GeV}, & J^P=2^+ ,\end{array} \right.\\
  m(X)&  = \left\{ \begin{array} {ll}4.07 {\rm GeV}, 4.12 {\rm GeV} , & J^P=0^+ , \\ 4.11 {\rm GeV} ,4.16 {\rm GeV} ,4.17 {\rm GeV} , & J^P=1^+ ,
 \\ 4.24 {\rm GeV}, & J^P=2^+ ,\end{array} \right.\\
 m(X')&  = \left\{ \begin{array} {ll}3.90 {\rm GeV}, 4.00 {\rm GeV} , & J^P=0^+ , \\ 3.93 {\rm GeV} ,4.02 {\rm GeV} , & J^P=1^+ ,
 \\ 4.09 {\rm GeV}, & J^P=2^+ .\end{array} \right.
\end{align}
Note that there are two tetraquarks have a very mass splitting around $4.02 {\rm GeV}$ for $X'$ with spin-parity $1^+$, and we only denote $4.02 {\rm GeV}$ once,
which shall be interpreted to two tetraquarks around $4.02 {\rm GeV}$.

Considering the orbitally excited tetraquarks with $L_{\delta \bar{\delta}'}=1$, the spin of tetraquark states can be 0, 1, 2, and 3. We particularly focus on  the $1^-$ tetraquark multiplet. We give the predictions for  their masses:
\begin{align}
 m(Z_c^{*}(1^-))&= \left\{ \begin{array} {ll}4.00 {\rm GeV}, 4.04 {\rm GeV} , 4.16 {\rm GeV} , \\ 4.22{\rm GeV} ,4.24 {\rm GeV},\end{array} \right.\\
 m(Z_{cs}^{*}(1^-))&= \left\{ \begin{array} {ll}4.27 {\rm GeV}, 4.31 {\rm GeV} , 4.37 {\rm GeV} , \\ 4.39{\rm GeV} ,4.44 {\rm GeV},4.45 {\rm GeV},\end{array} \right.\\
 m(X^{*}(1^-))&= \left\{ \begin{array} {ll}4.35 {\rm GeV}, 4.39 {\rm GeV} , 4.44 {\rm GeV} , \\ 4.46{\rm GeV} ,4.51 {\rm GeV},4.53 {\rm GeV},\end{array} \right.\\
 m(X^{'*}(1^-))&= \left\{ \begin{array} {ll}4.17 {\rm GeV}, 4.21 {\rm GeV} , 4.30 {\rm GeV} , \\ 4.31{\rm GeV} ,4.36 {\rm GeV},4.38 {\rm GeV}.\end{array} \right.\label{mass2}
\end{align}
Note that there are two tetraquarks have a very mass splitting around $4.16 {\rm GeV}$ for $Z_c^{*}$ with spin-parity $1^-$, and we  only denote $4.16 {\rm GeV}$ once

 There are a lot of  XYZ states discovered by different experiments.
Naive quark model where each meson is made of a quark-antiquark pair while each baryon is made of three quarks is not enough to explain these exotic structures. From Eq. (\ref{mass1}) to (\ref{mass2}), the mass spectra of hidden charm tetraquarks have been predicted, which may be used to explain
the XYZ states.

In Tabs.~\ref{spectra1} and \ref{spectra2}, we present these exotic XYZ states and also give  tetraquark interpretations. We find that a tetraquark $Z_c(3870)$ with $J^P=1^+$ may be used to explain the
$Z_c(3900)$ state,  while the radial excitation of tetraquark $Z_c(3870)$ may be used to explain the
$Z(4430)$ state. $Z_c(4020)$ may be treated as a companion of  $Z_c(3900)$, which is another  $J^P=1^+$ tetraquark from spin-spin splitting
interactions. Since $Z_c$, $Z_{cs}$, and $X$ form an octet representation in flavor SU(3) symmetry, it is reasonable to expect the existence of
$Z_{cs}$ states. However, there are no strong sign currently for the $J\psi K$ resonances. The best channels to hunting for $Z_{cs}$ are the processes with final
states $\psi(nS)+K^++K^-$, which will improve the significance and ovoid the interference
from other kinds of resonances except the $KK$ resonances. One may pay attention to
the invariant mass distribution of $J\psi K$ in the decay $B_c^+ \to J/\psi+K^++K^-+\pi^+$
measured by the LHCb Collaboration~\cite{Aaij:2013gxa}, where the data have a different shape compared to the theoretical predictions and may indicate new $J\psi K$ resonances. A comprehensive analysis in both theoretical and experimental sides is needed for a more determinate conclusion.

\begin{table}[thb]
\caption{\label{spectra1} Interpretation of the spectra (MeV in unit) of  the exotic X/Z states   as the possible hidden charm tetraquarks with positive parity. }
\begin{center}

\begin{tabular}{cccc}
\hline\hline
States ~~& Exp.  &  ~~States (Theo.) & $J^P$ (Theo.) \\
\hline
X(3823) &$3823.1\pm 1.9$\cite{Bhardwaj:2013rmw}& $X'(3900)$ & $0^+$   \\
X(3872) &$3871.68\pm 0.17$\cite{Choi:2003ue}& $X'(3930)$ & $1^+$  \\
$Z_c(3900)$ &$3891.2\pm 3.3$\cite{Ablikim:2013mio} & $Z_c(3870)$ & $1^+$  \\
X(3940) & $3942_{-8}^{+9}$\cite{Abe:2007sya} & $X'(4000)$ & $0^+$  \\
$Z_c(4020)$ &$4022.9\pm 2.8$\cite{Ablikim:2013wzq} & $Z_c(3880)$ & $1^+$  \\
X(4140) &$4156_{-25}^{+29}$\cite{Aaltonen:2009tz}
 & $X(4160)$ & $1^+$  \\
X(4274) &$4293\pm20$\cite{Aaltonen:2011at}
 & $X(4170)$ & $1^+$  \\
$Z_2(4250)$ &$4248_{-45}^{+185}$\cite{Mizuk:2008me}
& $Z_c(3720)(2S)$ & $0^+$  \\
$Z(4430)$ &$4458\pm15$\cite{Choi:2007wga}& $Z_c(3870)(2S)$ & $1^+$  \\
X(4500) &$4506\pm11$\cite{Aaij:2016iza} & $X'(3900)(2S)$  & $0^+$  \\
X(4700)&$4704\pm10$\cite{Aaij:2016iza}& $X(4120)(2S)$  & $0^+$  \\
\hline\hline
\end{tabular}
\end{center}
\end{table}

\begin{table}[thb]
\caption{\label{spectra2} Interpretation of the spectra (MeV in unit) of  the exotic X/Y states   as the possible hidden charm tetraquarks with negative parity. }
\begin{center}

\begin{tabular}{cccc}
\hline\hline
States ~~& Exp.  &  ~~States (Theo.) & $J^P$ (Theo.) \\
\hline
Y(4008) &$4008_{-49}^{+121}$\cite{Yuan:2007sj} & $Z_c^{*0}(4000)$ & $1^{--}$   \\
Y(4260) &$4263_{-9}^{+8}$ \cite{Aubert:2005rm} & $Z_c^{*0}(4240)$ & $1^{--}$   \\
Y(4274) &$4293\pm 20$\cite{Aaltonen:2011at} & $X^{'*}(4300)$ & $1^{-+}$   \\
Y(4360) & $4361\pm13$ \cite{Aubert:2007zz}& $X^{'*}(4360)$ & $1^{--}$   \\
X(4630) &$4634_{-11}^{+9}$\cite{Pakhlova:2008vn}
 & $X^*(4510)$ & $1^{--}$   \\
Y(4660) &$4664\pm 12$\cite{Wang:2007ea}
 & $X^*(4530)$ & $1^{--}$   \\
\hline\hline
\end{tabular}
\end{center}
\end{table}

$X(4140)$ and $X(4274)$ may be explained as $J^P=1^+$ tetraquarks with quark content $\frac{1}{\sqrt 6} (u\bar u+d \bar d -2s \bar s)c\bar c$.
$X(4500)$ may belong to flavor singlet in SU(3) symmetry, which can be explained as the radial excitation of $J^P=0^+$ tetraquarks with quark content $\frac{1}{\sqrt 3} (u\bar u+d \bar d +s \bar s)c\bar c$.
$X(4700)$ may be explained as the radial excitation of $J^P=0^+$ tetraquarks with quark content $\frac{1}{\sqrt 6} (u\bar u+d \bar d -2s \bar s)c\bar c$. $Y(4260)$ may be treated as $Z_c^{*0}(4240)$ tetraquark with $J^P=1^{--}$.

\section{Tetraquark two-body decays to charmed mesons\label{IV}}

In the previous section, the XYZ exotic states around charmonium region observed by different experiments
have been well classified into hidden charm octet or singlet tetraquarks. According to the possible interpretations of the XYZ states in Tabs. \ref{spectra1} and \ref{spectra2}, one can easily find that the predicted spin-parities are
in agreement with the experimental data, and most of their masses are very close to the experimental values.
In this section, we will phenomenologically discuss the tetraquark's decay modes, which will provide more internal information of the  observed  XYZ states, and also provide some channels to hunting for other  hidden charm tetraquarks.

The hidden charm tetraquarks lie above the $D\bar{D}$ threshold, thus they can decay into two charmed mesons.  The XYZ tetraquarks are denoted as $T_{c\bar{c}}$ for convenience. For the tetraquarks with positive parity, we have the two-body decay amplitudes
\begin{align}
{\cal M}(T_{c\bar{c}}[0^+] \to D \bar{D})=& m_{T_{c\bar{c}}} F_{T_{c\bar{c}}[0^+] D\bar{D}},\\
{\cal M}(T_{c\bar{c}}[0^+] \to D^* \bar{D}^*)=&  \varepsilon_{D^*,\mu}\varepsilon_{\bar{D}^*,\nu}(g^{\mu\nu}-\frac{P^\mu_{\bar{D}^*}P^\nu_{D^*}}{P_{D^*}\cdot P_{\bar{D}^*}})\nonumber\\&m_{T_{c\bar{c}}}F_{T_{c\bar{c}}[0^+]D^*\bar{D}^*},\\
{\cal M}(T_{c\bar{c}}[1^+] \to D \bar{D}^*)= &\varepsilon_{T_{c\bar{c}},\mu}\varepsilon_{\bar{D}^*,\nu}(g^{\mu\nu}-\frac{P^\mu_{\bar{D}^*}P^\nu_{T_{c\bar{c}}}}{P_{T_{c\bar{c}}}\cdot P_{\bar{D}^*}})\nonumber\\&\frac{m_{T_{c\bar{c}}}F_{T_{c\bar{c}}[1^+]D\bar{D}^*}}{\sqrt{3}},\\
{\cal M}(T_{c\bar{c}}[2^+] \to D^* \bar{D}^*)= &  (-\frac{P^\beta_{\bar{D}^*}P^\nu_{D^*}\varepsilon^{\mu\beta}_{T_{c\bar{c}}}}{P_{D^*}\cdot P_{\bar{D}^*}}-\frac{P^\mu_{\bar{D}^*}P^\alpha_{D^*}\varepsilon^{\alpha\nu}_{T_{c\bar{c}}}}{P_{D^*}\cdot P_{\bar{D}^*}}\nonumber\\&+\frac{P^\mu_{\bar{D}^*}P^\nu_{D^*}\varepsilon^{\alpha\beta}_{T_{c\bar{c}}}P^\beta_{\bar{D}^*}P^\alpha_{D^*}}{\left(P_{D^*}\cdot P_{\bar{D}^*}\right)^2}+\varepsilon^{\mu\nu}_{T_{c\bar{c}}})\nonumber\\&\varepsilon_{D^*,\mu}\varepsilon_{\bar{D}^*,\nu} \frac{m_{T_{c\bar{c}}}F_{T_{c\bar{c}}[2^+]D^*\bar{D}^*}}{\sqrt{5}},
\end{align}
where $F_{T_{c\bar{c}}D(D^*)\bar{D}(\bar{D}^*)}$ denotes the effective coupling to the tetraquark.

For the tetraquarks with $J^P=1^{--}$, we have the two body  decay amplitudes
\begin{align}
{\cal M}(T_{c\bar{c}}[1^{--}] \to D \bar{D})= & \varepsilon_{T_{c\bar{c}}} \cdot (P_D -P_{\bar{D}}) \frac{F_{T_{c\bar{c}}[1^{--}]D\bar{D}}}{\sqrt{3}},\\
{\cal M}(T_{c\bar{c}}[1^{--}] \to D \bar{D}^*)=& \varepsilon^{\mu}_{T_{c\bar{c}}} \varepsilon^{\nu}_{\bar{D}^*} P^\rho_D P^\sigma_{\bar{D}} \epsilon_{\mu\nu\rho\sigma}\nonumber\\
&\frac{F_{T_{c\bar{c}}[1^{--}]D\bar{D}^*}}{\sqrt{3}m_{T_{c\bar{c}}} },\\
{\cal M}(T_{c\bar{c}}[1^{--}] \to D^* \bar{D}^*)= & \varepsilon_{T_{c\bar{c}},\mu}\varepsilon_{\bar{D}^*,\rho}\varepsilon_{D^*,\nu}
\frac{F_{T_{c\bar{c}}[1^{--}]D^*\bar{D}^*}}{\sqrt{3}}\nonumber\\
&(g^{\mu\rho}(-P_{T_{c\bar{c}}}-P_{\bar{D}^*})^\nu+g^{\mu\nu}P^\rho_{T_{c\bar{c}}}, \nonumber\\
&+g^{\mu\nu}P^\rho_{D^*}+g^{\rho\nu}(P_{\bar{D}^*}-P_{D^*})^\mu).
\end{align}

The decay width of $T_{c\bar{c}} \to D(D^*)\bar{D}(\bar{D}^*)$ can be written as:
\begin{eqnarray}
\Gamma(T_{c\bar{c}} \to D(D^*)+\bar{D}(\bar{D}^*))&=&\frac{|\textbf{p}|}{8\pi
m_{T_{c\bar{c}}}^2}|{\cal M}|^2,
\end{eqnarray}
where
\begin{eqnarray}
|\textbf{p}|&=&\frac{\sqrt{\left(m_{T_{c\bar{c}}}^2-\left(m_1-m_{2}\right){}^2\right) \left(m_{T_{c\bar{c}}}^2-\left(m_{1}+m_{2}\right){}^2\right)}}{2 m_{T_{c\bar{c}}}},\nonumber
\end{eqnarray}
is the momentum modulus of final charmed meson in the tetraquark rest frame, and $m_1$ and $m_2$
are the masses of the final charmed mesons.

The corresponding ratios for the tetraquarks with positive parity   are
\begin{eqnarray}
\frac{\Gamma(T_{c\bar{c}}[0^+] \to D \bar{D})}{F_{T_{c\bar{c}}[0^+]D\bar{D}}^2|\textbf{p}|}&=&\frac{1}{8\pi
},\nonumber\\
\frac{\Gamma(T_{c\bar{c}}[0^+] \to D^* \bar{D}^*)}{F_{T_{c\bar{c}}[0^+] D^* \bar{D}^*}^2|\textbf{p}|}&=&\frac{3m_{T_{c\bar{c}}}^4+
8m_{T_{c\bar{c}}}^2|\textbf{p}|^2+48|\textbf{p}|^4}{8\pi
(m_{T_{c\bar{c}}}^2+4|\textbf{p}|^2)^2},\nonumber\\
\frac{\Gamma(T_{c\bar{c}}[1^+] \to D \bar{D}^*)}{F_{T_{c\bar{c}}[1^+] D \bar{D}^*}^2|\textbf{p}|}&=&\frac{h_1}{12\pi
(m_{T_{c\bar{c}}}^2+m_{D^*}^2-m_D^2)^2},\nonumber\\
\frac{\Gamma(T_{c\bar{c}}[2^+] \to D^* \bar{D}^*)}{F_{ T_{c\bar{c}}[2^+]D^* \bar{D}^*}^2|\textbf{p}|}&=&\frac{m_{T_{c\bar{c}}}^8}{\pi
(m_{T_{c\bar{c}}}^2+4|\textbf{p}|^2)^2}(\frac{1}{8}+\frac{|\textbf{p}|^2 h_2}{15m_{T_{c\bar{c}}}^8}),\nonumber\\
\end{eqnarray}
where
\begin{eqnarray}
h_1&=&m_{T_{c\bar{c}}}^4+4m_{T_{c\bar{c}}}^2m_{D^*}^2+m_{D^*}^4
\nonumber\\&&-2m_D^2(m_{D^*}^2+m_{T_{c\bar{c}}}^2)+m_D^4,\nonumber\\
h_2&=&15m_{T_{c\bar{c}}}^6+76m_{T_{c\bar{c}}}^4|\textbf{p}|^2+176m_{T_{c\bar{c}}}^2|\textbf{p}|^4+224|\textbf{p}|^6.\nonumber\\
\end{eqnarray}

The similar ratios for the tetraquarks with $J^P=1^{--}$ are
\begin{eqnarray}
\frac{\Gamma(T_{c\bar{c}}[1^{--}] \to D \bar{D})}{F_{T_{c\bar{c}}[1^{--}]D\bar{D}}^2|\textbf{p}|^3}&=&\frac{1}{6\pi
m_{T_{c\bar{c}}}^2},\nonumber\\
\frac{\Gamma(T_{c\bar{c}}[1^{--}] \to D \bar{D}^*)}{F_{ T_{c\bar{c}}[1^{--}]D \bar{D}^*}^2|\textbf{p}^3}&=&\frac{1}{12\pi
m_{T_{c\bar{c}}}^2},\nonumber\\
\frac{\Gamma(T_{c\bar{c}}[1^{--}] \to D^* \bar{D}^*)}{F_{T_{c\bar{c}}[1^{--}] D^* \bar{D}^*}^2|\textbf{p}|^3}&=&\frac{m_{T_{c\bar{c}}}^4-
\frac{104}{9}m_{T_{c\bar{c}}}^2|\textbf{p}|^2+\frac{48}{9}|\textbf{p}|^4}{2\pi
m_{T_{c\bar{c}}}^2(m_{T_{c\bar{c}}}^2-4|\textbf{p}|^2)^2}.\nonumber\\
\end{eqnarray}

One then easily find that  the decay widths of $1^{--}$ tetraquarks to double charmed mesons is suppressed compared to
that of the tetraquarks with positive parity to double charmed mesons. Take $D\bar{D}$ final states as an example,   the suppression factor is
\begin{eqnarray}
\frac{\Gamma(T_{c\bar{c}}[1^{--}] \to D \bar{D})}{\Gamma(T_{c\bar{c}}[0^+] \to D \bar{D})}&\simeq &\frac{4F_{ T_{c\bar{c}}[1^{--}]D \bar{D}}^2|\textbf{p}|^2}{3F_{ T_{c\bar{c}}[0^{+}]D \bar{D}}^2 m_{T_{c\bar{c}}}^2},
\end{eqnarray}

The value of  the first derivative of the radial wave function of the tetraquark at the origin can also be extracted from their leptonic pair decay
\begin{eqnarray}
 |R'_{\delta\delta'}(0)|^2= \frac{m_{Y_i}^4 \Gamma(T_{c\bar{c}}[1^{--}]\to \ell^+\ell^-)}{24\alpha^2 Q_i^2 },
\end{eqnarray}
where $|R'_{\delta\delta'}(0)|$ is the first  derivative of the radial wave function at the origin, $Q_i$ is defined as the effective charge of the diquark, where $Q_{[cu]}=4/3$ and $Q_{[cd]}=Q_{[cs]}=1/3$.

\section{Conclusion}
Dozens of the XYZ  mesons have been discovered by different experiments up to date, which call for
a deep understanding of the QCD spectrum. In this paper, we have studied the hidden charm octet and singlet tetraquarks in a diquark-antidiquark model. The mass spectra of hidden charm tetraquarks with different spin-parities  are given. According to the flavor SU(3) symmetry, most of XYZ states may be well-organized into octet and singlet representations.  The spin-parities and masses of the XYZ are predicted, most of which are in agreement with the data. We obtain a general tetraquark description for the XYZ states, i.e. octet and singlet representations. In order to improve
the calculation accuracy, a full fit including all the hadron spectra data to the parameters is needed, which we leave in future studies.

We find that $Z(4430)$ may be treated as the radial excitation of $Z_c(3900)$, and $Z_c(4020)$ may be treated as a companion of  $Z_c(3900)$, which is another  $J^P=1^+$ tetraquark from spin-spin splitting
interactions. $Z_{cs}$ as  the  octet representation companions of $Z_c(3900)$, and it is reasonable to expect their existences.
$X(4500)$ and  $X(4700)$ may be explained as the radial excitation of $J^P=0^+$ tetraquark, where $X(4500)$ is a flavor singlet with quark content $\frac{1}{\sqrt 3} (u\bar u+d \bar d +s \bar s)c\bar c$, while $X(4700)$ with  quark content $\frac{1}{\sqrt 6} (u\bar u+d \bar d -2s \bar s)c\bar c$. $Y(4260)$ may be treated as $Z_c^{*0}(4240)$ tetraquark with $J^P=1^{--}$.

Besides, the two-body hadronic decays of tetraquarks have been investigated, and we find that the decay widths of $1^{--}$ tetraquarks to double charmed mesons is suppressed compared to that of the tetraquarks with positive parity to double charmed mesons. Some relations among
different XYZ states are established in the tetraquark scheme, which can be checked in future experiments.

\section*{Acknowledgments}
This work was supported in part by the National Natural Science Foundation
of China under Grant No. 11235005, by a key laboratory grant from the Office of Science and Technology,
Shanghai Municipal Government (No. 11DZ2260700),
by Shanghai Natural  Science Foundation  under Grant No.15ZR1423100.

\end{document}